\documentclass[aps,prb,10pt,twocolumn]{revtex4-1}
\usepackage[ascii]{inputenc}
\usepackage{amsmath,amssymb,amsfonts,amsthm}
\usepackage{graphicx}
\usepackage{bbm}
\usepackage[caption=false]{subfig}
\usepackage[pdftex,bookmarks=false,colorlinks=true,linkcolor=blue,
citecolor=blue,filecolor=black,urlcolor=blue]{hyperref}

\newcommand{\ud}{\mathrm{d}}

\newcommand{\R}{\mathbb{R}}

\DeclareMathOperator{\e}{e}
\DeclareMathOperator{\tr}{tr}

\graphicspath{{figures/}}

\begin{document}

\title{Time evolution of matrix product operators with energy conservation}

\author{Christian B.~Mendl}
\email{christian.mendl@tu-dresden.de}
\affiliation{Technische Universit\"at Dresden, Institute of Scientific Computing, Zellescher Weg 12-14, 01069 Dresden, Germany}

\date{December 31, 2018}

\begin{abstract}
We devise a numerical scheme for the time evolution of matrix product operators by adapting the time-dependent variational principle for matrix product states [J.~Haegeman et~al, Phys.~Rev.~B 94, 165116 (2016)]. A simple augmentation of the initial operator $\mathcal{O}$ by the Hamiltonian $H$ helps to conserve the average energy $\tr[H \mathcal{O}(t)]$ in the numerical scheme and increases the overall precision. As demonstration, we apply the improved method to a random operator on a small one-dimensional lattice, using the spin-1 Heisenberg XXZ model Hamiltonian; we observe that the augmentation reduces the trace-distance to the numerically exact time-evolved operator by a factor of $10$, at the same computational cost.
\end{abstract}

\maketitle

\section{Introduction}

The real-time evolution of strongly correlated quantum systems poses a fundamental and computationally challenging task. Recent interest has been spurred by the question of how quantum information spreads in correlated systems, characterized by out-of-time-ordered correlation functions\cite{ShenkerStanford2014, Kitaev2014, Hosur2016, RakovszkyPRX2018}. Another recent approach are hydrodynamic descriptions\cite{SpohnNonlinearHydro2014, MendlSpohn2013, BhaseenNatPhys2015, BertiniPRL2016, CastroAlvaredoPRX2016, HolographicQuantumMatter2018}, based on local conservation laws (in one dimension) of the form
\begin{equation}
\label{eq:microscopic_conservation_law}
\frac{\ud}{\ud t} Q_n(t) + \mathcal{J}_{n+1}(t) - \mathcal{J}_n(t) = 0.
\end{equation}
Here $n$ is the lattice site index, $Q_n$ is a conserved field operator (like particle number or spin, or local energy) and $\mathcal{J}_n$ denotes the corresponding local current. Taking thermal averages $\langle \cdot \rangle \equiv \frac{1}{Z} \tr[\e^{-\beta H} \cdot]$ and assuming slow variation on the scale of the lattice, Eq.~\eqref{eq:microscopic_conservation_law} transforms into the Euler equation $\partial_t q(x, t) + \partial_x \mathsf{j}(x,t) = 0$, where $x \in \R$ is the continuum version of $n$.

Abstractly, let $\mathcal{O}$ be a linear operator (not necessarily Hermitian) acting on the quantum Hilbert space. Our goal is to approximate the time-evolved operator $\mathcal{O}(t) = \e^{i H t} \mathcal{O} \e^{-i H t}$ (with $H$ the Hamiltonian) by numerically solving the corresponding Heisenberg equation of motion
\begin{equation}
\label{eq:heisenberg}
\frac{\ud}{\ud t} \mathcal{O}(t) = i [H, \mathcal{O}(t)]
\end{equation}
while preserving the average energy $\tr[H \mathcal{O}(t)]$ in the numerical scheme.

For simulating quasi one-dimensional quantum systems on classical computers, density matrix renormalization group (DMRG) methods and modern formulations within the matrix product state (MPS) framework \cite{WhiteDMRG1992, Schollwock2011} have emerged as one of the most successful methods. The time-dependent variational principle (TDVP) for matrix product states is the canonical approach for solving the Schr\"odinger equation $i \frac{\ud}{\ud t} \lvert\Psi\rangle = H \lvert\Psi\rangle$ for a wavefunction $\lvert\Psi\rangle$ projected onto the MPS manifold of given bond dimensions\cite{Haegeman2016}. The desirable properties of the one-site integration scheme in Ref.~\onlinecite{Haegeman2016} include the exact conservation of the norm $\langle \Psi \vert \Psi \rangle^{1/2}$ and the average energy $\langle \Psi \vert H \vert \Psi \rangle$. In this work, we adapt the TDVP method to matrix product operators (MPO), and augment the initial operator $\mathcal{O}$ to improve the numerical accuracy.

\section{Interpreting operators as states}

\begin{figure}[b]
\centering
\includegraphics[width=0.75\columnwidth]{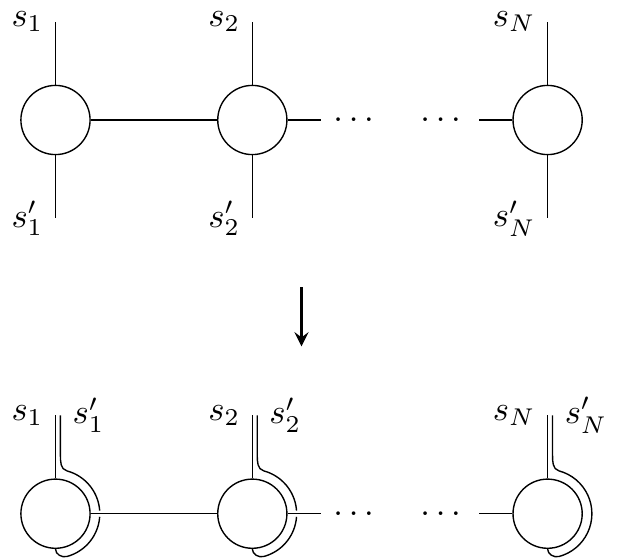}
\caption{Interpreting a matrix product operator as a matrix product state.}
\label{fig:mpo_as_mps}
\end{figure}

We assume that $\mathcal{O}$ is a MPO acting on a lattice with $N$ sites (using open boundary conditions) of the form
\begin{equation}
\label{eq:O_mpo_def}
\mathcal{O}[A] = \sum_{s,s'} A^{s_1,s_1'}(1) A^{s_2,s_2'}(2) \cdots A^{s_N,s_N'}(N) \, \lvert s \rangle \langle s' \rvert.
\end{equation}
Here each $A^{s_n,s_n'}(n)$ is a site-dependent complex matrix of dimension $D_{n-1} \times D_{n}$ (with $D_0 = 1$ and $D_N = 1$), and the components of the indices $s_n$ and $s_n'$ run from $1$ to $d$, with $d$ the local Hilbert space dimension. To cast the time evolution into the form of a Schr\"odinger equation and render it amendable to the TDVP scheme for \emph{states}, our first step is a ``purification'' of $\mathcal{O}$ (see also Refs.~\onlinecite{VerstraetePRL2004, ZwolakVidal2004, FeiguinWhite2005}); in our case combining the two physical dimensions per site into one large dimension $d^2$, as depicted in Fig.~\ref{fig:mpo_as_mps}.
We denote the resulting state by $\lvert\mathcal{O}\rangle$; its MPS representation literally agrees with \eqref{eq:O_mpo_def} after grouping indices as $\sigma_n = (s_n,s_n')$ and formally replacing $\lvert s \rangle \langle s' \rvert$ by $\lvert \sigma \rangle$. Consequently, the inner product of two linear operators $\mathcal{O}_1$ and $\mathcal{O}_2$ equals
\begin{equation}
\langle \mathcal{O}_1 \vert \mathcal{O}_2 \rangle = \tr[\mathcal{O}_1^{\dagger}\,\mathcal{O}_2].
\end{equation}

With the definition $\hat{H} = -[H, \cdot] \equiv - H \otimes \mathbbm{1} + \mathbbm{1} \otimes H^T$ (where the second tensor factor corresponds to the primed indices), we can now formally express Eq.~\eqref{eq:heisenberg} as Schr\"odinger equation
\begin{equation}
\label{eq:schroedinger}
i \frac{\ud}{\ud t} \lvert\mathcal{O}\rangle = \hat{H} \lvert\mathcal{O}\rangle.
\end{equation}
Note that, by construction, $\hat{H} \lvert H \rangle = 0$.

\section{Modified TDVP method}

The TDVP method approximates Eq.~\eqref{eq:schroedinger} by projecting the evolution vector onto the MPS manifold (for fixed virtual bond dimensions) at the current state\cite{Haegeman2016}:
\begin{equation}
\label{eq:tdvp_ode}
\frac{\ud}{\ud t} \lvert\mathcal{O}[A]\rangle = -i \hat{P}_{\mathcal{T}_{\lvert\mathcal{O}[A]\rangle}} \hat{H} \lvert\mathcal{O}[A]\rangle.
\end{equation}
The explicit form of the projector $\hat{P}_{\mathcal{T}_{\lvert\mathcal{O}[A]\rangle}}$ has been derived in Ref.~\onlinecite{Haegeman2016} and forms the basis for a Lie-Trotter splitting scheme. Interestingly, this scheme preserves norm and energy (defined via $\hat{H}$) exactly. In our case this ``energy'' equals
\begin{equation}
\langle\mathcal{O}\rvert \hat{H} \lvert\mathcal{O}\rangle = - \tr\left[ \mathcal{O}^{\dagger} H \mathcal{O} - \mathcal{O} H \mathcal{O}^{\dagger} \right],
\end{equation}
which unfortunately differs from the physical energy $\tr[H \mathcal{O}]$. As $H$ is Hermitian, $\tr[H \mathcal{O}] = \langle H \vert \mathcal{O} \rangle$, and the physical energy conservation may be interpreted as $\frac{\ud}{\ud t} \mathcal{O}(t)$ being perpendicular to $\lvert H \rangle$. This property is (in general) not satisfied exactly by Eq.~\eqref{eq:tdvp_ode} since $\hat{P}_{\mathcal{T}_{\lvert\mathcal{O}[A]\rangle}} \lvert H \rangle \neq \lvert H \rangle$.

We can ameliorate this issue by using that (i) any MPS $\lvert\Psi[A]\rangle$ is contained in its tangent space, i.e., $\hat{P}_{\mathcal{T}_{\lvert\Psi[A]\rangle}} \lvert\Psi[A]\rangle = \lvert\Psi[A]\rangle$, and (ii) $\e^{-i \hat{H} t} \lvert H \rangle = \lvert H \rangle$ since $\hat{H} \lvert H \rangle = 0$. Specifically, we start from the initial MPS
\begin{equation}
\label{eq:X_init}
\lvert\mathcal{X}[\tilde{A}]\rangle = \lvert\mathcal{O}[A]\rangle + \gamma \lvert H \rangle
\end{equation}
(with a small parameter $\gamma \ll 1$) and then apply the TDVP time evolution to $\lvert\mathcal{X}[\tilde{A}]\rangle$:
\begin{equation}
\label{eq:tdvp_extend_ode}
\frac{\ud}{\ud t} \lvert\mathcal{X}[\tilde{A}]\rangle = -i \hat{P}_{\mathcal{T}_{\lvert\mathcal{X}[\tilde{A}]\rangle}} \hat{H} \lvert\mathcal{X}[\tilde{A}]\rangle.
\end{equation}
The time-evolved state $\lvert\mathcal{O}(t)\rangle$ is then approximated as
\begin{equation}
\label{eq:tdvp_extend_result}
\lvert\tilde{\mathcal{O}}(t)\rangle = \lvert\mathcal{X}[\tilde{A}(t)]\rangle - \gamma \lvert H \rangle.
\end{equation}

The summation in \eqref{eq:X_init} can be evaluated exactly using MPS techniques (via block-diagonal tensors, effectively summing respective virtual bond dimensions). Since the virtual bond dimensions of a typical quantum Hamiltonian (on quasi one-dimensional lattices) is relatively small, the computational cost like\-wise increases only moderately when employing \eqref{eq:tdvp_extend_ode} instead of \eqref{eq:tdvp_ode}. For the direct comparison in the following Sect.~\ref{sec:example_heisenberg}, we will actually use the same maximal bond dimensions during the TDVP time evolution.

It turns out that $\hat{P}_{\mathcal{T}_{\lvert\mathcal{X}[\tilde{A}]\rangle}} \lvert H \rangle = \lvert H \rangle$ holds exactly at the initial $\lvert \mathcal{X}[\tilde{A}] \rangle$, which follows from the block-diagonal structure of the MPS tensors representing the sum of the two states in \eqref{eq:X_init}. Thus, evaluated at the initial $\lvert \mathcal{X}[\tilde{A}] \rangle$:
\begin{equation}
\label{eq:X_deriv_perp_H}
\frac{\ud}{\ud t} \langle H \vert \mathcal{X}[\tilde{A}] \rangle = -i \langle H \rvert \hat{P}_{\mathcal{T}_{\lvert\mathcal{X}[\tilde{A}]\rangle}} \hat{H} \lvert\mathcal{X}[\tilde{A}]\rangle = 0.
\end{equation}
One expects that this relation still holds approximately during the TDVP time evolution since $\e^{-i \hat{H} t} \lvert\mathcal{X}[\tilde{A}]\rangle = \e^{-i \hat{H} t} \lvert\mathcal{O}[A]\rangle + \gamma \lvert H \rangle$, i.e., the form of \eqref{eq:X_init} is preserved by the exact time evolution.

The $\gamma$ parameter ensures that $\lvert\mathcal{X}[\tilde{A}]\rangle$ is close to $\lvert\mathcal{O}[A]\rangle$, which we have found advantageous to increase the numerical precision of the time evolution. On the other hand, we observe that the resulting state $\lvert\tilde{\mathcal{O}}(t)\rangle$ only depends weakly on the precise value of $\gamma$, as expected, such that a fine-tuning of $\gamma$ is not required.

\section{Application to the spin-1 XXZ Heisenberg model}
\label{sec:example_heisenberg}

To demonstrate and benchmark the numerical scheme, we consider the (non-integrable) spin-1 Heisenberg XXZ chain with Hamiltonian (setting $\hbar = 1$)
\begin{equation}
H = J \sum_n \left( \tfrac{1}{2} \big(S^{+}_{n} S^{-}_{n+1} + S^{-}_{n} S^{+}_{n+1}\big) + \Delta S^z_{n} S^z_{n+1} \right)
\end{equation}
where $S^{\pm}_n = S^{x}_n \pm i S^{y}_n$ and $S^{x}_n, S^{y}_n, S^{z}_n$ are the usual spin-1 operators with eigenvalues $\{-1, 0, 1\}$ acting on lattice site $n$. The local Hilbert space dimension is thus $d = 3$. The time-evolved spin operator $S^{z}_n(t) = \e^{i H t} S^{z}_n \e^{-i H t}$ obeys the microscopic conservation law
\begin{equation}
\label{eq:Sz_conservation_law}
\frac{\ud}{\ud t} S^{z}_n(t) + \mathcal{J}^{z}_{n+1}(t) - \mathcal{J}^{z}_n(t) = 0
\end{equation}
with the local spin current
\begin{equation}
\mathcal{J}^{z}_n = J \left( S^{x}_{n-1} S^{y}_{n} - S^{y}_{n-1} S^{x}_{n} \right),
\end{equation}
which follows from a straightforward evaluation of the commutator $i [H, S^{z}_n]$. From Eq.~\eqref{eq:Sz_conservation_law} one concludes that $\sum_n S^{z}_n$ is conserved in time (assuming periodic boundary conditions). Similarly, the local energy operator
\begin{equation}
h_n = J \left( \tfrac{1}{2} \big(S^{+}_{n} S^{-}_{n+1} + S^{-}_{n} S^{+}_{n+1}\big) + \Delta S^z_{n} S^z_{n+1} \right)
\end{equation}
obeys the microscopic conservation law
\begin{equation}
\label{eq:energy_conservation_law}
\frac{\ud}{\ud t} h_n(t) + \mathcal{J}^{\epsilon}_{n+1}(t) - \mathcal{J}^{\epsilon}_n(t) = 0
\end{equation}
with the energy current
\begin{equation}
\begin{split}
\mathcal{J}^{\epsilon}_n(t) &= J_x J_y \left( S^y_{n-1} S^z_{n} S^x_{n+1} - S^x_{n-1} S^z_{n} S^y_{n+1} \right) \\
& + \text{cyclic permutations of } (x, y, z)
\end{split}
\end{equation}
and $J_x = J_y = J$, $J_z = J \Delta$.

We consider a small system with $N = 6$ lattice sites and open boundary conditions, such that exact diagonalization is feasible for obtaining reference quantities. The Hamiltonian parameters are chosen as $J = 1$ and $\Delta = 1.2$. The nonzero complex entries of the tensors $A(n)$ of the initial operator $\mathcal{O}[A]$ are independently drawn from the standard normal distribution, separately for real and imaginary parts. We choose the small initial bond dimension $2$, but zero-pad the tensors $A(n)$ to accommodate a maximum bond dimension of $81$, since the one-site TDVP method\cite{Haegeman2016} used for the time evolution leaves the bond dimensions invariant. Explicitly, the virtual bond dimensions $D_n$ are $(1, 9, 81, 81, 81, 9, 1)$. As last step of the initialization, the tensors $A(n)$ are left-normalized by QR decompositions. Due to the sparsity pattern of the initial $A(n)$ tensors, we can represent the sum in Eq.~\eqref{eq:X_init} without increasing the bond dimension. The tensors representing $\lvert H \rangle$ are temporarily scaled by $10^{-3}$ before adding it to $\lvert\mathcal{O}[A]\rangle$, thus $\gamma = 10^{-3 N} = 10^{-18}$. The simulations are performed using the PyTeNet software package\cite{PyTeNet}.

\begin{figure}[!ht]
\centering
\includegraphics[width=\columnwidth]{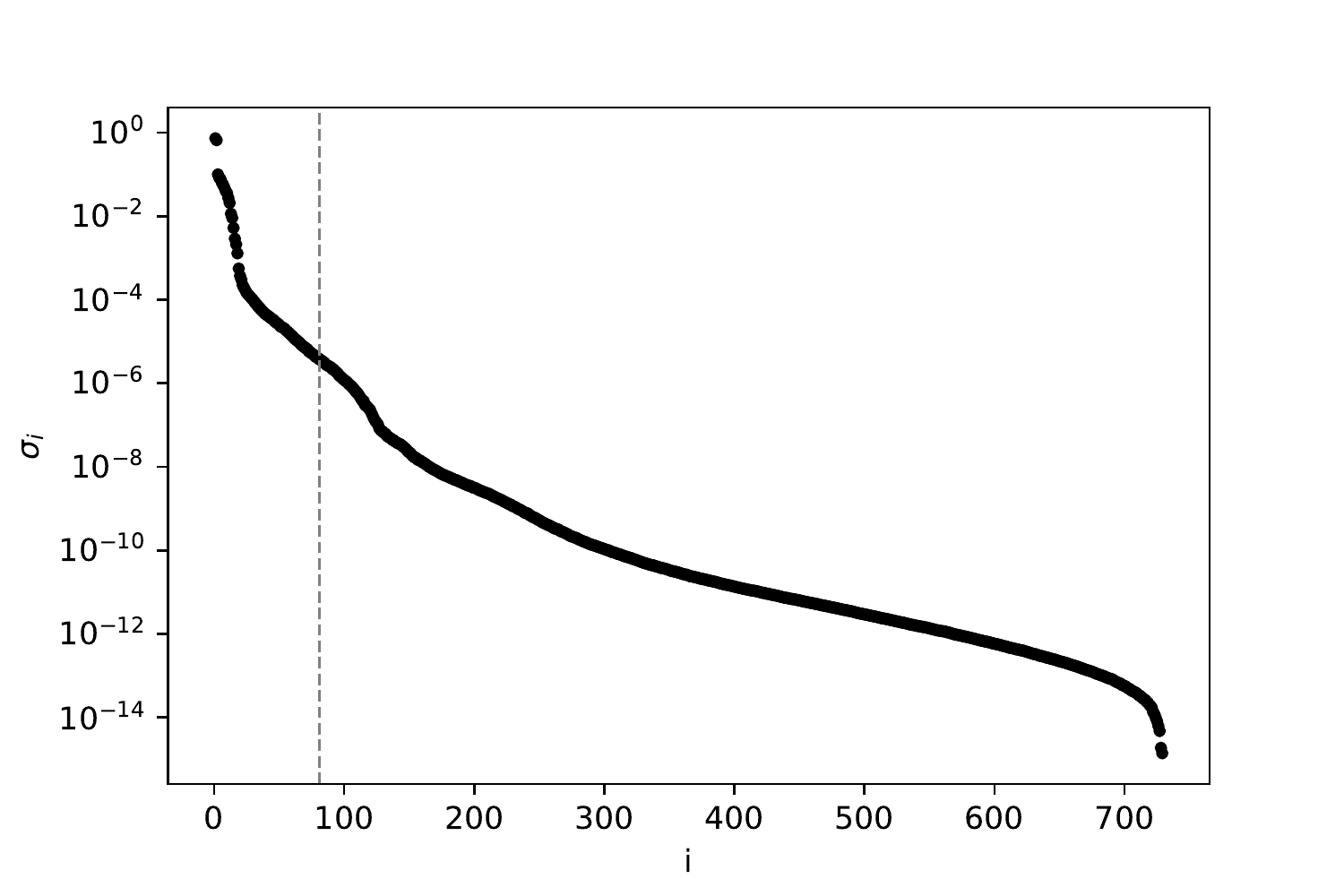}
\caption{Schmidt coefficients of the exact $\mathcal{O}(t)$ at $t = \frac{1}{8}$, for symmetric left-right partitioning of the lattice ($N = 6$ sites). The dashed vertical line marks the $81$-th coefficient.}
\label{fig:schmidt_coefficients}
\end{figure}

We time-evolve up to $t = \frac{1}{8}$ to ensure that the Schmidt coefficients of $\mathcal{O}(t)$ (partitioning into left and right halves with $N/2$ sites) still decay fast, i.e., the exact $\mathcal{O}(t)$ can in principle be well approximated by a matrix product operator of maximal bond dimension $81$, see Fig.~\ref{fig:schmidt_coefficients}. The corresponding von Neumann entanglement entropy is $1.287$ at $t = \frac{1}{8}$. For comparison, it reads $0.9913$ at $t = 0$.

\begin{figure}[!ht]
\centering
\includegraphics[width=\columnwidth]{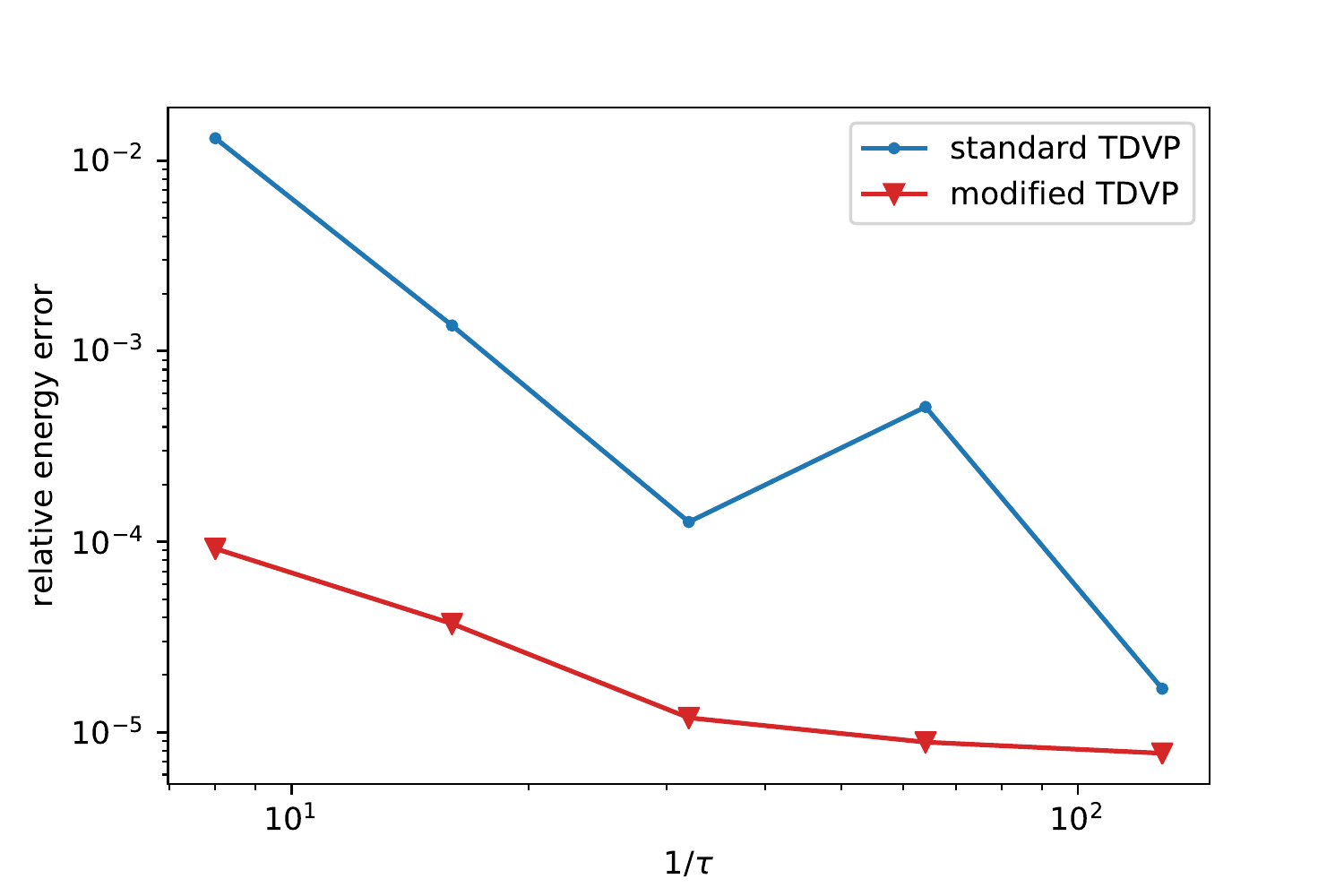}
\caption{Relative error of the average energy $\tr[H \mathcal{O}]$ at $t = \frac{1}{8}$, as a function of the time step $\tau$. The blue curve corresponds to the conventional TDVP method of Eq.~\eqref{eq:tdvp_ode} (one-site integration scheme)\cite{Haegeman2016}, compared to the modified version of Eqs.~\eqref{eq:X_init}, \eqref{eq:tdvp_extend_ode} and \eqref{eq:tdvp_extend_result} shown in red. In both cases the maximal virtual bond dimension during the time evolution is $81$.}
\label{fig:energy_error}
\end{figure}

We now evaluate the accuracy of the standard TDVP method \eqref{eq:tdvp_ode} and the modified version \eqref{eq:X_init}, \eqref{eq:tdvp_extend_ode}, \eqref{eq:tdvp_extend_result}, first in terms of energy conservation, i.e., the deviation of $\tr[H \mathcal{O}]$ from its initial value. Fig.~\ref{fig:energy_error} shows that, indeed, the energy error is appreciably smaller for the modified method, as expected based on Eq.~\eqref{eq:X_deriv_perp_H}.

\begin{figure}[!ht]
\centering
\includegraphics[width=\columnwidth]{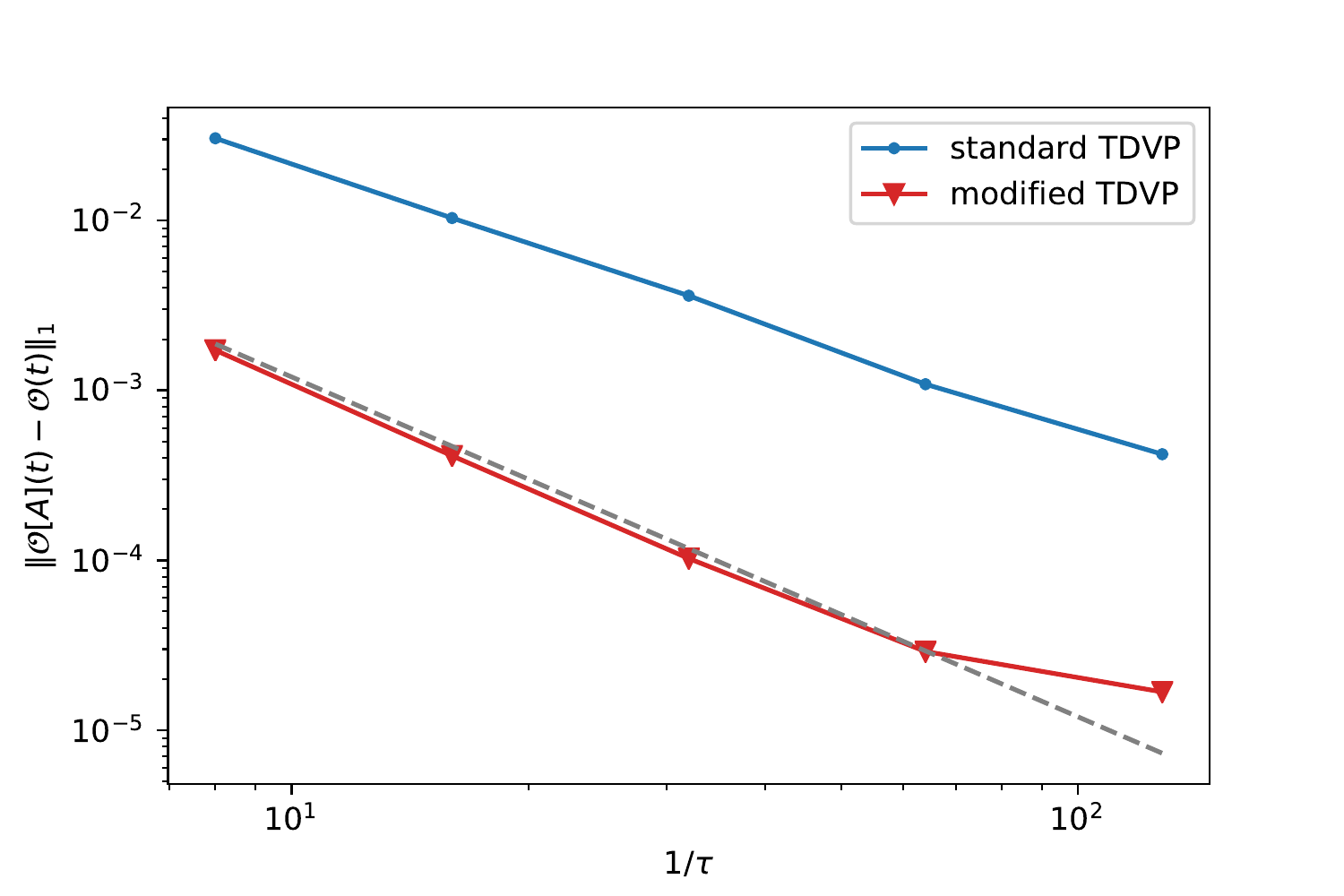}
\caption{Convergence rate with respect to time step size $\tau$ quantified via the trace-distance from the numerically exact operator $\mathcal{O}(t)$, for the same simulation as in Fig.~\ref{fig:energy_error}. The gray dashed line visualizes the expected $\sim \tau^2$ scaling.}
\label{fig:operator_error}
\end{figure}

We quantify the overall accuracy via the trace-distance (matrix $1$-norm) of $\mathcal{O}[A]$ and $\tilde{\mathcal{O}}(t)$ from the numerically exact $\mathcal{O}(t)$, respectively, see Fig.~\ref{fig:operator_error}. Interestingly, the modified TDVP method reduces the error by a factor of $10$, for the same bond dimensions used during the time evolution.

\section{Discussion and outlook}

While we have focused on energy conservation, the proposed modification of the TDVP method in Eqs.~\eqref{eq:X_init}, \eqref{eq:tdvp_extend_ode} and \eqref{eq:tdvp_extend_result} works for any operator $\mathcal{K}$ commuting with the Hamiltonian, since the corresponding conservation law can be written as
\begin{equation}
\frac{\ud}{\ud t} \langle \mathcal{K}^{\dagger} \vert \mathcal{O}(t) \rangle = \frac{\ud}{\ud t} \tr[\mathcal{K}\, \mathcal{O}(t)] = \tr[\mathcal{K}\, i [H, \mathcal{O}(t)]] = 0.
\end{equation}
Moreover, it is feasible to take several conservation laws simultaneously into account, by forming the sum
\begin{equation}
\lvert\mathcal{X}[\tilde{A}]\rangle = \lvert\mathcal{O}[A]\rangle + \sum_j \gamma_j \lvert \mathcal{K}_j \rangle.
\end{equation}
The effect of such a modification on the overall accuracy is an interesting question for future studies.

Considering practical implementations of the TDVP method for operators, we remark that the Hamiltonian $\hat{H} = -[H, \cdot] \equiv - H \otimes \mathbbm{1} + \mathbbm{1} \otimes H^T$ can be applied without explicitly forming the outer Kronecker product with identity matrices, in order to improve the computational efficiency.


\end{document}